# Multiple subglacial water bodies below the South Pole of Mars unveiled by new MARSIS data

Sebastian Emanuel Lauro, Dipartimento di Matematica e Fisica, Università degli studi Roma Tre, Rome, Italy, sebastian.lauro@uniroma3.it

Elena Pettinelli*, Dipartimento di Matematica e Fisica, Università degli studi Roma Tre, Rome, Italy elena.pettinelli@uniroma3.it

Graziella Caprarelli, School of Sciences, University of Southern Queensland, Toowomba, Australia Graziella.Caprarelli@usq.edu.au

Luca Guallini, Istituto di Radioastronomia (IRA), Istituto Nazionale di Astrofisica (INAF), Bologna, Italy, luca.guallini@inaf.it

Angelo Pio Rossi, Department of Physics and Earth Sciences, Jacobs University Bremen, Bremen, Germany, an.rossi@jacobs-university.de

Elisabetta Mattei, Dipartimento di Matematica e Fisica, Università degli studi Roma Tre, Rome, Italy elisabetta.mattei@uniroma3.it

Barbara Cosciotti, Dipartimento di Matematica e Fisica, Università degli studi Roma Tre, Rome, Italy barbara.cosciotti@uniroma3.it

Andrea Cicchetti, Istituto di Astrofisica e Planetologia Spaziali (IAPS), Istituto Nazionale di Astrofisica (INAF), Rome, Italy, andrea.cicchetti@inaf.it

Francesco Soldovieri, Istituto per il Rilevamento Elettromagnetico dell'Ambiente, Consiglio Nazionale delle Ricerche, Naples, Italy , soldovieri.f@irea.cnr.it

M. Cartacci, Istituto di Astrofisica e Planetologia Spaziali (IAPS), Istituto Nazionale di Astrofisica (INAF), Rome, Italy, marco.cartacci@inaf.it

F. Di Paolo, Dipartimento di Matematica e Fisica, Università degli studi Roma Tre, Rome, Italy dipaolo@fis.uniroma3.it

R. Noschese, Istituto di Astrofisica e Planetologia Spaziali (IAPS), Istituto Nazionale di Astrofisica (INAF), Rome, Italy, raffaella.noschese@inaf.it

R. Orosei, Istituto per il Rilevamento Elettromagnetico dell'Ambiente, Consiglio Nazionale delle Ricerche, Naples, Italy, roberto.orosei@inaf.it

**The detection of liquid water by the Mars Radar for Subsurface and Ionospheric Sounding (MARSIS), at the base of the South Polar Layered Deposits (SPLD) in Ultimi Scopuli, has**

**reinvigorated the debate about the origin and stability of liquid water under present-day Martian conditions. To establish the extent of subglacial water in this region, we acquired new data, achieving unprecedented radar coverage over the study area. Here we present and discuss the results obtained by a new method of analysis of the complete MARSIS dataset, based on signal processing procedures usually applied to terrestrial polar ice sheets. Our results strengthen the claim of the detection of a liquid water body at Ultimi Scopuli and indicate the presence of other wet areas nearby. We suggest that the waters are hypersaline perchlorate brines, known to form at Martian polar regions and shown to survive for geologically significant periods of time at below-eutectic temperatures.**

Mars' geological record, albeit incomplete, indicates that the climate has undergone dramatic changes throughout its history. The transfer of water between the Martian cryolithospheric and atmospheric global reservoirs is linked to orbital dynamics[1]. It is generally accepted that quasi-periodic variations of orbital eccentricity[2] and planetary obliquity[3-6] profoundly affected the Martian climate[7]. Proof of the effects of orbital forcing on the climate of Mars is found in the stratigraphy of the North Polar Layered Deposits (NPLDs), where there is good correspondence between the timescales of layer deposition and astronomical cycles[5]. The relationship between orbital parameters and the actual evolution of the layered deposits at the south pole is however still largely unconstrained[8,9]. Under present-day climate conditions, Martian polar caps are generally assumed to be cold-based[10]. However, the Late Noachian-Early Hesperian circumpolar Dorsa Argentea Formation displays characteristic esker-like morphologies that have been interpreted as evidence of basal melting under thick ice sheets[11], possibly caused by a combination of warmer and wetter past climate conditions[12] and a significantly higher heat flux (~ 45-60 mW/m$^2$) than at present (20-22 mW/m$^2$)[13]. Furthermore, km-scale tectonic deformations in the Late Amazonian South Polar Layered Deposits (SPLDs)[14,15] provide evidence of broad ice movement, suggesting localized basal melting of the Martian southern ice sheet under possible warmer conditions[16].

The set of physical conditions conducive to basal melting at the Martian polar and subpolar regions has been explored[16-18] via the theoretical combination of heat flow parameters that could result in ice melting at the base of the south polar ice cap. Recently Sori and Bramson[19] suggested that a high geothermal gradient (heat flux ≥72 mW/m$^2$) is needed for basal melting of the SPLDs, regardless of the salinity level. The same authors thus postulated that magmatic activity must have occurred in the region less than 1 Myr ago for a liquid water body to exist at the base of the SPLDs. Evidence for late Amazonian (as young as 2 Myr) magmatic activity has been reported in the Elysium region[20,21], suggesting the potential for localized high geothermal gradients on Mars in recent geological times. An anomalously high geothermal gradient is, however, not the only possible cause for temperature increases at the base of extensive ice-sheets. For example, localized basal melting of a thick Early Amazonian polythermal ice sheet in Isidis Planitia was modeled by Souček et al.[22], who concluded that subglacial wet areas could form under climatic and geologic conditions that are not significantly different from those on present-day Mars.

On Earth, one of the most valuable methods to detect subglacial liquid water is the Radio-Echo Sounding (RES)[23] technique, which employs bursts of radio waves to image buried geological structures in a similar fashion to active seismic prospection. As radio waves propagate with little attenuation in ice, this method is particularly well suited to study the internal structure of the Martian Polar Layer Deposits (PLDs)[10] and to detect the bedrock below such deposits[24]. The possible identification of subsurface liquid water was one of the main goals behind the development of the Mars Advanced Radar for Subsurface and Ionosphere Sounding (MARSIS). The sounder was launched in 2003 onboard the Mars Express spacecraft and began to collect data in the summer of 2005. After several years of data acquisition, however, the lack of any clear evidence of basal liquid water below the Martian Polar Caps seemed to indicate that water, if present, may be located at a greater depth than previously thought[25,26]. Recently, however, radar detection of subglacial liquid water in Ultimi Scopuli[27], at the base of the SPLDs, reignited the scientific debate about present-day stability of liquid water at the Martian polar regions. Because the nature of the body of liquid water was not addressed in detail in the previous study, our purpose here is to better constrain and define the characteristics and spatial distribution of the subglacial bright areas associated with basal liquid water and to discuss possible physical and chemical conditions that can explain the formation and persistence of such water at the Martian south polar regions.

We analyzed MARSIS observations applying a methodological approach adapted from signal processing procedures commonly used in terrestrial radar sounding studies to discriminate between wet and dry subglacial basal conditions. In Antarctica[23] and Greenland[28], the combination of qualitative (bedrock morphology in the radar image) and quantitative (signal features) analysis is used to detect subglacial water. East Antarctica was the first area where subglacial lakes were unequivocally found by RES investigations and where specific criteria for lake detection and classification were proposed[29] (Methods). Such criteria are suitable for well-defined large subglacial lakes but fail in the detection of more complex water distribution[30]. In Greenland, given the paucity of subglacial lakes[31], the criteria tested in East Antarctica have found very limited application. In this region RES data have been mainly used to constrain the subglacial basal conditions and to define the spatial distribution of the water (ponded/thawed/frozen) at the ice-sheet bed. Recently, however, several previously undetected subglacial lakes have been found by applying some of the criteria originally used in East Antarctica[32]. Some of these criteria have also been applied to RES data collected in the Canadian Arctic, resulting in the discovery of the first two isolated hypersaline subglacial lakes on Earth[33].

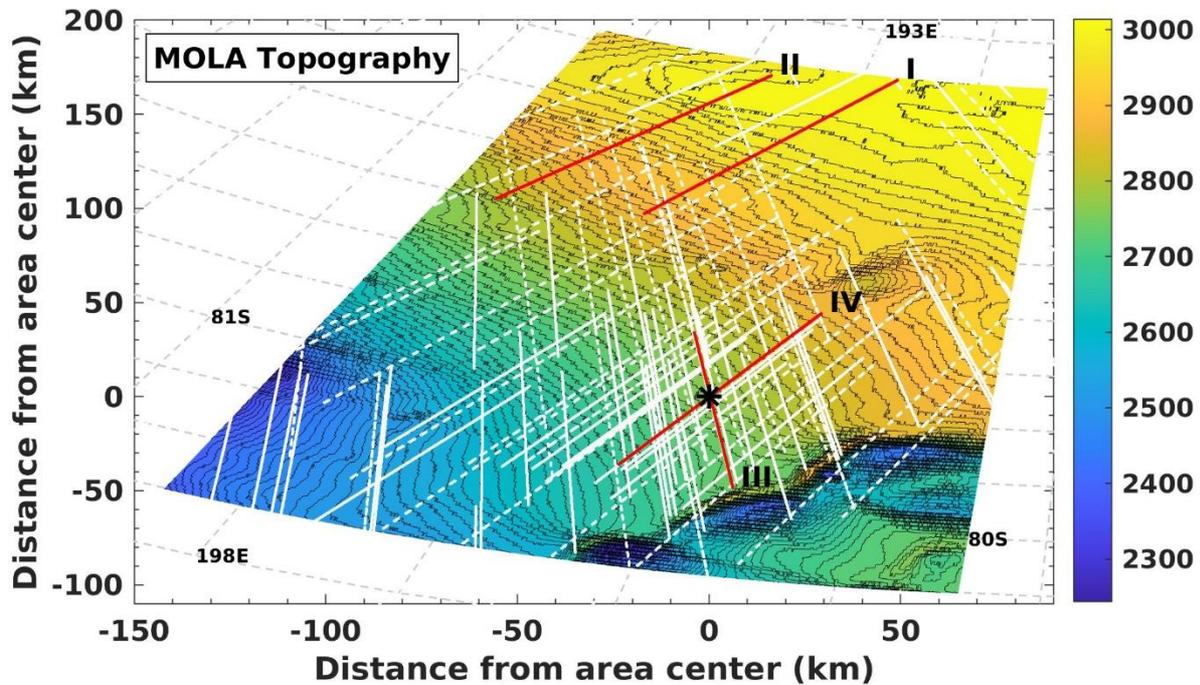

Fig. 1 Ultimi Scopuli Mars Orbiter Laser Altimeter (MOLA) topographic map and location of MARSIS radar profiles collected in the region. Solid white lines are Super-frame data and dash white lines are Flash memory data (Methods). Red lines (observation number I, II, III, IV) highlight the radar profiles used to describe the three diagnostic parameters (Fig.3) that were applied to discriminate between wet and dry areas. The black star indicates the center of the bright area detected by Orosei et al. [27] at 193°E and 81°S.

MARSIS technical characteristics severely limit the application of most of the criteria used in East Antarctica (Methods); conversely, the methodology tested in Greenland to discriminate between ponded water and frozen or dry basal conditions is fully applicable to MARSIS data. Such a methodology is based on signal intensity (which is an indication of the basal reflectivity), signal acuity (which is associated with the smoothness of the bed[34]) and bed-echo intensity variability (which detects the transition from dry to wet materials at the base of the ice[35]) (Methods).

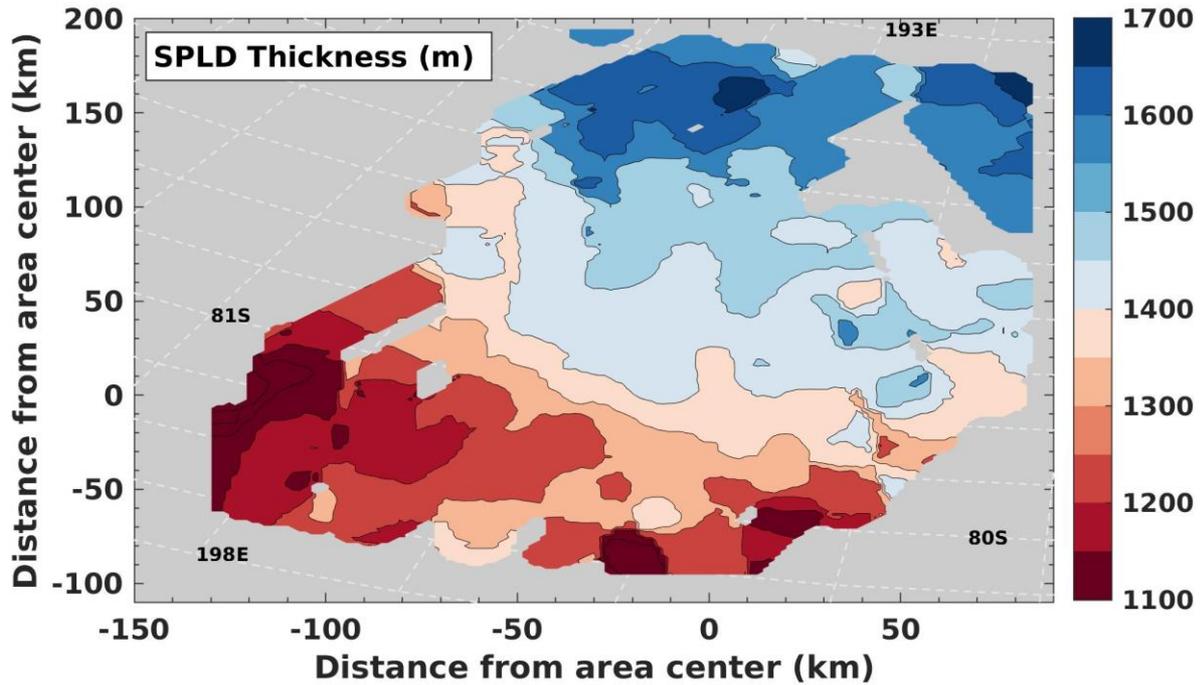

Fig. 2 SPLD thickness computed assuming a velocity value of 163 m/μs ($\varepsilon_{ice}$=3.4) which corresponds to water ice with 10% admixed dust. Gray areas indicate no data available.

The analysis presented in this work was performed on the data acquired in 134 radar profiles during multiple campaigns over Ultimi Scopuli from 2010 to 2019 (Fig. 1). We focused on a 250x300 km² zone, around and including the bright area previously identified by Orosei et al.[27], where we have significant data coverage (90 observations) and many radar profiles that cross each other. Given the MARSIS pulse length of about 200 m in air, the surface of the studied area can be considered smooth with elevation gently decreasing northward (Fig. 1). The basal interface (bottom of the SPLDs) is clearly detectable in all radar profiles and it is thus possible to estimate the local thickness of the SPLDs under the assumption that the signal velocity does not change in the entire investigated area. Immediately around the bright area the thickness of the SPLDs is constant (considering the pulse length and the footprint) but at the regional scale it progressively decreases in the same direction as the surface (Fig. 2), resulting in an essentially flat basal topography (Supplementary Fig.2). In this work we present and discuss only the data collected at 4MHz, as it is the largest and most robust dataset (Methods).

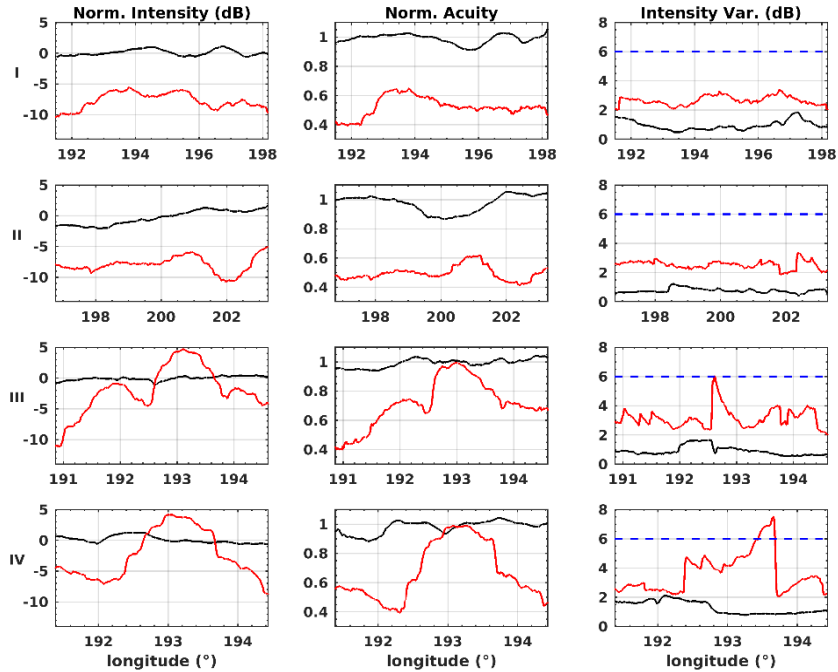

Fig. 3 Data collected outside and inside the bright area. From top to bottom observations are: I, II, III, IV. Black lines indicate surface and red lines basal parameters. In the right column the dashed blue line indicates the water detection threshold (6 dB) according to Jordan et al.[35].

In order to describe the meaning of the three chosen diagnostic parameters (intensity, acuity and bed-echo intensity variability) in terms of basal water detection, we selected four representative radar profiles in the investigated region (red lines in Fig. 1): two outside (observations I and II) and two across (observations III and IV) the previously detected bright area. We analyzed the trend of these parameters along the observations and compared the spatial behavior of each parameter between observations (see Fig. 3). The surface values of intensity and acuity (black lines in Fig. 3) are broadly similar in both areas with limited variations along each observation, indicating a smooth and flat surface at the MARSIS wavelength. Conversely, the basal values (red lines in plots of Fig. 3) are markedly different. In the background area, along orbits I and II, basal intensity and acuity values are constant and much lower than the corresponding surface values, whereas the intensity variability is always above the surface values. These parameters suggest a low reflectivity of the basal material (-10 dB relative to the surface), a relatively rough basal interface (low acuity) and a spatially homogeneous bedrock (relatively constant intensity variability along tracks). Conversely, the observations acquired across the bright area show a marked increase of the basal intensity (about 10 dB) along track, reaching a maximum value (well above the surface values) at the center of the bright area. Similar trends are observed in the basal acuity values, while intensity variability values change abruptly where the observations approach the bright area. According to Oswald et al.[34], the occurrence of high intensity and high acuity values in the same location indicates the presence of ponded water, and Jordan et al. [35] have shown that intensity variability values exceeding 6 dB mark the transition (edge detector) between dry and wet materials. Moreover, Dowdeswell and Siegert[36] emphasized that a change in basal intensity of about 10 dB along track could be evidence of the presence of a lake, whereas smaller variations (e.g., on the order of 2dB) could indicate wet sediments or water intruded in the bedrock around a lake[29]. The combination of the three criteria makes the interpretation of the basal conditions along the four profiles quite robust and suggests a remarkable difference in the bed material properties between the two areas: observations I and II detected a dry (or frozen) bedrock, whereas observations III and IV crossed at least one large water ponded area.

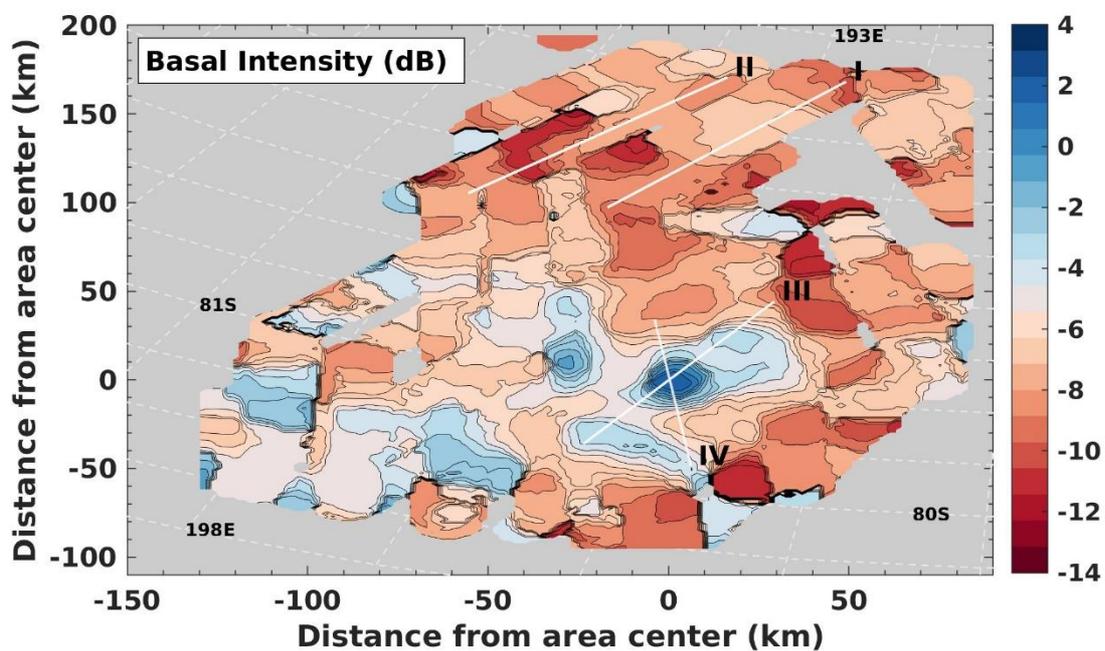

a)

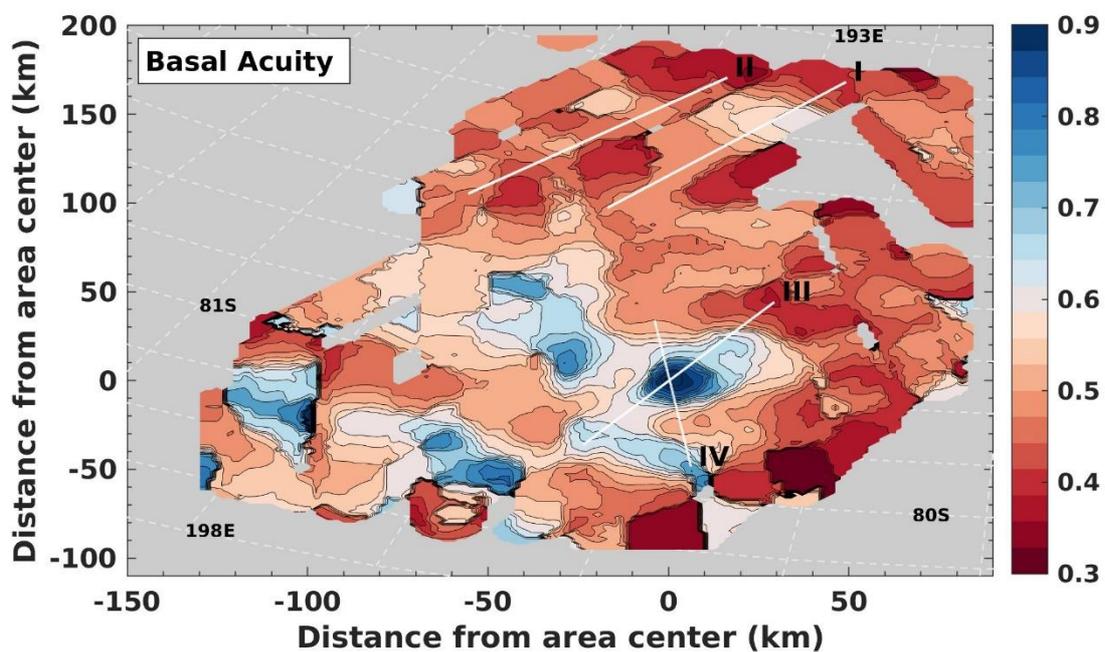

b)

Fig. 4 Spatial distribution of the normalized basal reflectivity (a) and normalized basal acuity (b) computed from the radar data collected at 4 MHz. White lines highlight the same red lines (observations) shown in Fig. 1 and used for the analysis of Fig. 3. Both maps clearly show two distinct areas: an upper southern area characterized by a very low and relatively constant signal intensity [from -14 to -6 dB] and acuity [from 0.3 to 0.6] and a lower northern area characterized by several patches of high signal intensity [from -3 to 4 dB] and acuity [from 0.65 to 1].

To assess the wet/dry spatial distribution of the basal material below the SPLDs, we used all observations collected in the study area (Fig. 1) and generated a basal intensity map and a basal acuity map (Fig. 4). Comparison of the two maps highlights the strong spatial correlation between the two parameters, with only a few exceptions where high values of basal acuity do not correspond to high values of basal intensity and vice versa. Therefore, following Oswald et al.[34], we conclude that the basal material in the southern area is uniformly dry whereas the northern area is characterized by the presence of several basal patches of ponded water.

These results corroborate the initial discovery by Orosei et al.[27] of a stable body of liquid water in Ultimi Scopuli using a different and independent technique, highlighting at the same time a more extensive, complex scenario with ubiquitous water patches surrounding the subglacial lake. This is illustrated considering the correlation of the features mapped in Fig. 4 and the plots of the radar parameters of orbit III (Fig. 3). The trend of basal intensity and acuity in the plots shows that the area centered at 193°E-81°S (point 0,0 in the maps) is characterized by a large body of water encircled by patchy water pools or wet areas of smaller extent. This interpretation is also supported by the intensity variability values computed on other observations partially crossing the lateral water patches (Supplementary Fig.4 and Fig.5). The abrupt transitions in bed material properties are still well detectable, however the variability values at the edge of the patches are slightly lower than the 6dB threshold[35], probably due to the fact that the patches are not completely intercepted by the radar footprint and/or that they consist of wet sediments or small volumes of water (Supplementary Fig.3).

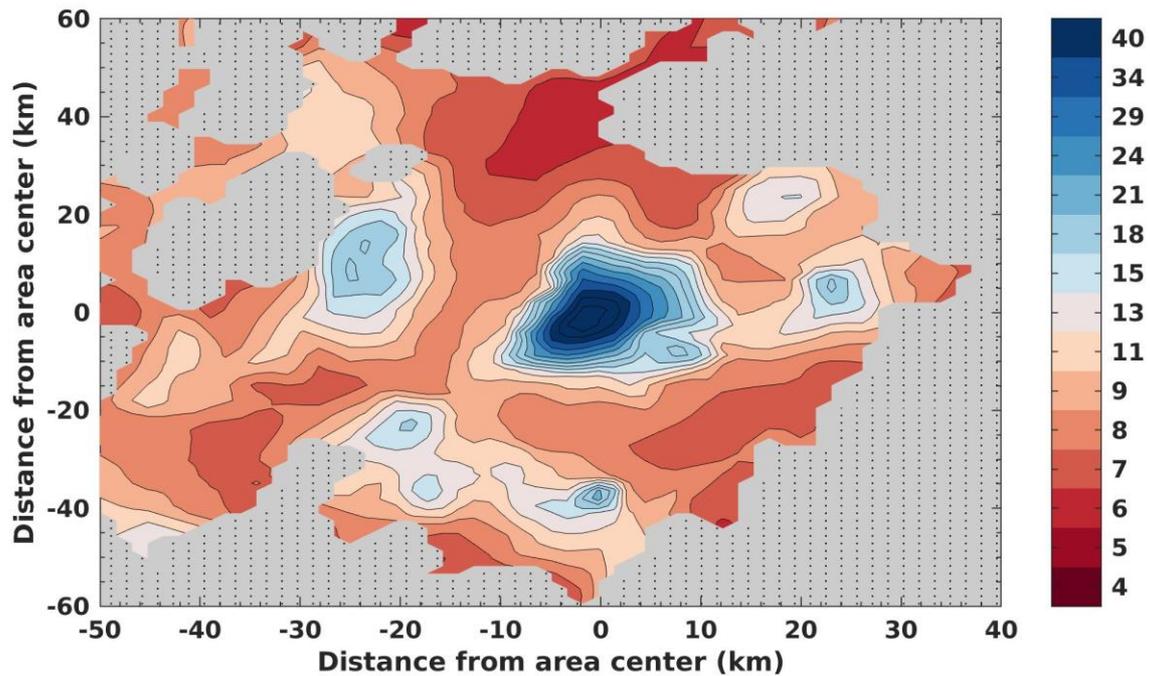

Fig. 5 Relative dielectric permittivity map computed by inverting the radar data considering all regions where the number of samples is larger than 100. The map only shows the permittivity values retrieved from radar data having acuity values larger than 0.5 (Methods). This procedure has reduced the dimension of the study area to 90x120 km$^2$. Values larger than 15 suggest the presence of liquid water.

As a final step, we applied the approach proposed by Lauro et al.[37] to compute the basal water distribution map in terms of permittivity (Fig.5), in the area where a significant number of samples are present (Supplementary Fig. 6) and where the radar profiles cross each other (Methods). The map shows that the areas with high values of basal permittivity (15-40) correspond to the smooth areas (high acuity values) shown in Fig.4b, whereas the surroundings exhibit much lower permittivity values (about 6-8). On Earth, a permittivity value of 15 can be considered a threshold, with higher values indicating the presence of liquid water in the basal material[7] and lower values indicating that the material is dry or frozen. Based on this consideration, the main body of water (with its average permittivity of ~40) has a dimension of approximately 20x30km, it probably contains the largest volume of water and it is separated from the other small water pools by strips of dry basal material.

The unique combination of physical, geological, climatic and topographic conditions that could favor the formation of liquid water and its long-term survival in a subglacial lake at the base of the SPLDs is, at best, a matter of speculation at this point in time. It is, however, possible to interpret relevant observations in the context of known experimental data and terrestrial analog studies to propose plausible processes for the formation and persistence of subglacial liquid water on Mars. Combining evidence from radar datasets and thermal models, subglacial hypersaline aqueous solutions were found to persist on Earth at temperatures much lower than the freezing point of water[38].

Thus, brines have been proposed as the most plausible form of liquid water on the Martian surface and subsurface under present-day physical conditions[39].

The process of absorption of atmospheric water by perchlorates and the subsequent formation of hypersaline solutions (i.e., deliquescence) was directly observed at the Phoenix Landing Site[40]. Considering that Ca-, Mg-, Na- and K- perchlorates, chlorates and hydrated chlorides[41-44] are globally ubiquitous in the Martian regolith, we posit that deliquescence and the formation of brines could plausibly occur at the south polar latitudes as well. Experimental work has shown that soluble salts with low eutectic temperatures deliquesce at low relative humidity values over a wide range of temperatures, overlapping with those expected on Mars[45-47], suggesting that brines may readily form in sub-polar regions when the temperatures are in the higher range (e.g., at noon). Re-crystallization of brines (efflorescence) when temperatures drop, however, is often kinetically inhibited[48] because high activation energies are required for the transition from liquid to solid (ordered) states. Freezing experiments conducted under conditions similar to those on Mars have shown that perchlorate and chloride brines may exist for long times after their formation without effloressing[49,50]. It is therefore plausible that once formed, brines may exist on Mars in a metastable state for geologically significant periods of time[51].

Orosei et al.[27] suggested that the subglacial water discovered at Ultimi Scopuli could be hypersaline solutions. Subsequently, Sori and Bramson [19] computed the geothermal flux at the base of the SPLDs that would melt ice when Na-, Mg- and Ca- perchlorates are present in the icy mixture. They used Pestova et al.'s [52] eutectics for Mg- and Ca-perchlorate aqueous solutions, and Chevrier et al.'s[53] eutectics for Na- perchlorate solutions, determining that an anomalously high geothermal flux of 72 mW/m$^2$ is required for the icy mixture to achieve the temperature of the lowest eutectic (Ca-perchlorate, 199 K). Recent experiments have shown, however, that Mg- and Ca-perchlorate-H$_2$O solutions remain liquid in a super-cooled state at temperatures as low as 150 K[54]. Mean temperatures at the Ultimi Scopuli location have been estimated to be approximately 160 K at the surface[27], increasing with depth by a few to a few tens of K per km, depending on the unknown geothermal flux and thermal properties of the SPLDs. These temperatures are very close to the lower boundary of super-cooled solutions, where kinetic processes are particularly important. We argue therefore that thermophysical modeling based on equilibrium conditions may not be wholly realistic in this context and propose instead that metastable conditions are likely to produce a geologically significant effect, both in terms of the formation of brines and in terms of their longevity on Mars.

Recently, Arnold et al. [55], used the subsurface echo time delay data presented in Orosei et al.[27] to estimate the SPLD basal topography around the area of the previously detected lake[27]. The authors used the resulting SPLD bed elevations to calculate the subglacial hydraulic potential surface, finding that the high reflectivity area did not correspond to any model-predicted lake location and concluding that the bright reflector should be a hydraulically isolated patch of liquid rather than a subglacial lake. This conclusion is based on a well-established methodology[56,57], but we note that it depends critically on the accuracy of the basal topography, as stated also in Arnold et al.[55]. Orosei et al.[27] cautioned that the large size of the MARSIS footprint and the diffuse nature of basal echoes outside the bright reflectors prevent a detailed and accurate reconstruction of the basal topography: simple averaging of echo time delays in overlapping footprints, which was used to estimate subglacial topography in both Orosei et al.[27] and Arnold et al.[55], cannot improve either horizontal or vertical resolution. By taking the vertical resolution of MARSIS (~55m in ice) as the nominal uncertainty on basal topography, it is found that the corresponding uncertainty in the hydraulic potential is about 1 MPa (Supplementary Fig.2), which is larger than the variation of the hydraulic potential across the area surrounding the bright reflector, affecting the reliability of the local minima in the hydraulic potential.

Our new methodological approach, applied to the expanded MARSIS data set, reinforces Orosei et al.'s[27] interpretation of the existence of a subglacial water body at Ultimi Scopuli and provides new evidence of the presence of other discrete wet areas nearby. It is thus reasonable to surmise that patches of ponded liquid water may be ubiquitous at the base of the SPLD, although it might prove impossible for MARSIS to detect them because surface morphologies at Ultimi Scopuli are uniquely suited to quantitative space-borne radar data analysis and such conditions may not be present elsewhere. It is our view that, in the absence of heat flow data or geological evidence pointing to geothermal anomalies, models advocating recent magmatic activity to explain melting at the base of the SPLD rely on largely speculative assumptions that disregard other key evidence acquired from planetary observations to date. We propose instead that known physical and chemical properties of hypersaline aqueous solutions provide a viable interpretive framework based on current observations and measurements of properties of the Martian surface and subsurface. We thus favor the hypothesis that SPLD basal liquid water exists because of a high concentration of dissolved salts, an interpretation consistent with our analysis of MARSIS data.

The possibility of extended hypersaline water bodies on Mars is particularly exciting because of the potential for the existence of microbial life, such as extremophiles, anaerobes[58] or even aerobes (considering that the solubility of $O_2$ in brines is up to 6 times the minimum level required for microbial respiration[59]). The water bodies at the base of the SPLD therefore represent areas of potential astrobiological interest and planetary protection concern, and future missions to Mars should target this region to acquire experimental data in relation to the basal hydrologic system, its chemistry, and traces of astrobiological activity.

**Acknowledgments**: This work was supported by the Italian Space Agency (ASI) through contract ASI-INAF 2019-21-HH.0. MARSIS is operating onboard the European Space Agency spacecraft Mars Express. We would like to thank S.E, Beaubien for careful proofreading of the manuscript and improvement of the English.

**Authors contributions:** S.E.L. and E.P. designed the research, developed the methodology, performed data analysis, and wrote the manuscript; G.C., L.G. and A.P.R. developed the geological context, and contributed to writing relevant sections of the manuscript; G.C. and E.P. developed and wrote the geochemical discussion; E.M. and B.C. performed data analysis, contributed text and figures, and co-wrote the manuscript; A.C. conducted data pre-processing; F.S. contributed to inverse electromagnetic modelling, to the interpretation of the geophysical data, and to writing of the manuscript; M.C. contributed to data acquisition and analysis, and discussed ideas; F.D.P. contributed text and figures and discussed ideas; R.N. contributed to data acquisition and analysis; R.O. discussed ideas and co-wrote the manuscript. S.E.L., E.P., G.C., L.G., A.P.R., E.M, B.C., F.S. and R.O. addressed reviewers' and editor's comments to all iterations of the manuscript.

**Competing interests**: The authors declare no competing interests.

**Methods**

**MARSIS data.** MARSIS radar is a nadir-looking pulse limited radar sounder that can operate in two main observational modalities: Sub-Surface (SS) Mode and Active Ionosphere Sounding (AIS) Mode. In SS mode, MARSIS transmits a 250 µs chirp with a 1 MHz bandwidth. According to the predicted Solar Zenith Angle (SZA), the chirp central frequency is selected among 4 different values (1.8, 3, 4 and 5 MHz) to work well above the cut-off plasma frequency of the Martian ionosphere[27]. After range compression and Hanning windowing, the achievable range resolution in pure ice is about 55 m (assuming a velocity of 170 m/µs). In this work, we only analyzed unprocessed data to avoid the uncertainty due to the incoherent integration performed on- board in normal mode[27]. Two alternative data acquisition methods were used[60]: i) the Flash Memory (FM) technique, which collects discontinuous intervals of unprocessed/raw data along orbits; and ii) the Superframe acquisition mode, which continuously collects data but along shorter orbits (Fig.1).

MARSIS data were collected in different years (2010 – 2019) and during different Martian seasons, therefore each orbit refers to particular conditions of the Martian ionosphere. Because the ionosphere can cause dispersion on the transmitted signal, reducing the echo intensity and producing a broadening of the received signal, we normalized all quantities (i.e., surface and basal intensity and acuity) to the median of the relevant surface quantity along each orbit. In particular, the use of the median minimizes the effects caused by local surface echo power fluctuations, which are sometimes observed in the data, without altering the along track variation of the basal reflectivity and acuity.

The 4 MHz data set was the most complete, and therefore we applied our analysis to this specific data set to improve the robustness of our statistical analysis. The data collected at the other frequencies are sparser (especially the 3 MHz, which is also more affected by the ionosphere) and thus less statistically significant (Supplementary Fig.1). Despite this fact, the analysis of such data supports the results obtained at 4 MHz.

**Radar signal features for subglacial water detection.** Historically, four specific criteria have been proposed to identify and categorize the subglacial lakes in East Antarctica[29]: i) standard deviation of the echo strength (values lower than 3 dB indicate that the basal interface is smooth at the scale of the radar footprint); ii) high echo strength relative to the immediate surroundings; iii) absolute echo strength (related to the basal reflection coefficient); and iv) the assumption that a subglacial lake is hydraulically flat[29]. According to such criteria East Antarctic subglacial lakes have been classified as: *definite lakes*, at least partially satisfying all four criteria; *dim lakes*, which satisfy the first two quantitative criteria; *fuzzy lakes*, which only satisfy the absolute and relative signal intensity requirements; *indistinct lakes*, characterized by low standard deviation in echo intensity; and *failing lakes*, that only satisfy one of the four criteria[29]. Recently, however, such criteria have been partially modified and updated to identify more complex subglacial water distribution (e.g., active lakes) by using the specularity content, which is related to the variations in the angular distribution of radar bed echo scattering[30,61].

MARSIS radar is, in principle, similar to the systems used in RES investigations, but several aspects prevent the application of most of the East Antarctica criteria[29]. Given the operating frequency (1.8 - 5 MHz), the antenna dimension (40 m) and the altitude of the spacecraft (250 - 900 km), MARSIS pulse limited radius footprint is very large (from 6 to 11 km) compared to footprints of the common RES radar allocated on airborne platforms and working at higher frequencies (typically of the order of 100 m). The bright area below the SPLDs interpreted as a stable body of liquid water[27] was estimated to be approximately 20 km in extent, i.e., comparable to the dimension of the MARSIS pulse limited footprint. For these reasons, it is not possible to compute the standard deviation of the echo strength (first criterion in Carter et al.[29]). Moreover, as the MARSIS antenna could not be calibrated[27] and the signal absorption in the SPLDs is not well constrained, the absolute echo strength (second criterion in Carter et al. [29]) cannot be applied. Finally, the vertical (~55 m in ice) and horizontal (~7 km) resolution of MARSIS prevents a detailed characterization of the bedrock morphology, topography and hydraulic potential (forth criterion in Carter et al. [29]), as the uncertainties associated with the estimation of these parameters are similar in magnitude to the range of measured variations (Supplementary Fig.2).

**Spatial smoothing and acuity.** We processed MARSIS data according to the method developed by Oswald and Gogineni[28], which applies an along-track average to the radar traces (spatial smoothing) in order to reduce the power variance due to variable roughness[28]. In our analysis, the along-track waveform averaging window $W$ is set equal to the diameter of the pulse-limited footprint area:

$$W = 2\sqrt{c\frac{p}{2}\left(H + \frac{z}{n}\right)} \cong 2\sqrt{c\frac{p}{2}H}, \qquad (1)$$

where $p = 1\ \mu s$ is the transmitted pulse length, $H$ is the spacecraft altitude (this quantity can vary between 250 and 900 km), $c$ is velocity of light in a vacuum, $z$ is the depth of the reflector and $n \cong 1.71$ is the water ice refractive index. For example, considering a spacecraft height $H = 400$ km the resulting window is $W \cong 15\ km$.

In addition, we computed the acuity $A_c$, which is an indication of interface roughness, from the depth aggregated echo power[34] as follows:

$$A_c = \frac{max\{x_r^2(t)\}}{\sum_{t=\tau-T/2}^{\tau+T/2} x_r^2(t)}, \qquad (2)$$

where $x_r(t)$ is the received pulse generated by a reflector located at a time delay $\tau$ (e.g., surface or bedrock), $T = \sqrt{\frac{H}{2vc}}$ ($v$ is the operating frequency).

**Basal topography and hydraulic potential.** We computed the basal topography ($B$) and the hydraulic potential ($\phi_H$) along the observations reported in Fig.3 according to the following equations[29]:

$$B = S - h \qquad (3)$$

and

$$\phi_H = \rho_w g_M (S - h(1 - \rho_{ice}/\rho_w)) \qquad (4)$$

Where $S$ is the surface topography (based on MOLA data), $h$ is the SPLDs thickness computed assuming a permittivity $\varepsilon_{ice} = 3.4$, $\rho_w = 1980 \text{ kg/m}^3$ is the density of the perchlorate solution[18], $\rho_{ice} = 1100 \text{ kg/m}^3$ the average density of the SPLDs[62] and $g_M = 3.72 \text{ m/s}^2$ is the Martian gravity (Supplementary Fig.2). The uncertainties $\Delta u$ associated with these parameters (where $u = B$ or $u = \phi_H$, depending on the used equation) have been computed applying the statistical propagation formula, under the assumption that all uncertainties are independent and uncorrelated:

$$\Delta u = \sqrt{\sum_i \left(\frac{\partial u}{\partial x_i} \Delta x_i\right)^2} \quad (5)$$

where $x_i$ are the variables (e.g. $S, h, ...$) in equations 3 and 4 and $\Delta x_i$ the associated uncertainties. In Eq. (5) $\Delta h = 61m$, $\Delta S = 61m$ and $\Delta \rho_{ice} = 115 \text{ kg/m}^3$. Note that we have neglected the uncertainties related to the density of perchlorate brines and Martian gravity.

**Bed-echo intensity variability.** We used the intensity variability parameter $\sigma_I$ to localize the transition between dry (or frozen) and wet bed conditions[35]. The parameter $\sigma_I$ is given by:

$$\sigma_I = \Delta R \sqrt{f^2(1-f) + (1-f)^2 f} \quad (6)$$

where

$$\Delta R = 20 \log_{10} \left( \left| \frac{\sqrt{\varepsilon_b} - \sqrt{\varepsilon_{ice}}}{\sqrt{\varepsilon_b} + \sqrt{\varepsilon_{ice}}} \right| \left| \frac{\sqrt{\varepsilon_{dry}} + \sqrt{\varepsilon_{ice}}}{\sqrt{\varepsilon_{dry}} - \sqrt{\varepsilon_{ice}}} \right| \right) \quad (7)$$

With $\varepsilon_b$ the dielectric permittivity of the basal material, $\varepsilon_{ice}$ the permittivity of the SPLDs and $\varepsilon_{dry}$ the permittivity of the dry rock. In addition, $f$ is the fraction of wet area (wet–dry mixing ratio in Jordan et al. [35]) inside the radar footprint. Supplementary Fig.3 illustrates the intensity variability as a function of the basal permittivity for three values of the mixing ratio $f$, assuming a threshold of 15 for the permittivity of wet materials[27].

We computed the intensity variability $\sigma_I$ (expressed in dB) along each orbit at $x_i$ position, as follows:

$$\sigma_I(x_i) = \frac{10}{\ln(10)} \frac{\sqrt{\frac{1}{N} \sum_{x=x_i-W/2}^{x_i+W/2} [P_{ag}(x) - <P_{ag}(x_i)>]^2}}{<P_{ag}(x_i)>} \quad (8)$$

where $P_{ag}(x)$ is the aggregated power computed according to Jordan et al. [35], and $<P_{ag}(x_i)>$ is the average echo power:

$$<P_{ag}(x_i)> = \frac{1}{N} \sum_{x=x_i-W/2}^{x_i+W/2} P_{ag}(x) \quad (9)$$

The observations collected around the bright area show specific values of the bed-echo intensity variability. In particular, the radar observations (A and D in Supplementary Fig.5) crossing approximately the center of the main body of liquid water (Fig.3 and Supplementary Fig.4 and Fig.5) exhibit an intensity variability exceeding the fairly conservative threshold of 6 dB[35] and therefore clearly indicate a transition from dry to wet basal material. On the other hand, the observations passing on the edge of the main body or on the other patches (B,C, E and F in Supplementary Fig.5), even if still showing an abrupt change in intensity variability, do not exceed 4 - 5 dB. These results can be explained considering the intersection between the radar footprint and the bed conditions (dry/wet), which is accounted for by the parameter *f*. Indeed, wet areas smaller than the radar footprint (i.e., having a low fraction of wet material illuminated by the radar antenna) cannot generate abrupt intensity variability as high as 6 dB, but can still indicate a transition in the basal conditions from dry to wet (Supplementary Fig.3, Fig.4 and Fig.5).

**Permittivity map.**

The basal permittivity map was generated by applying an inversion probabilistic approach to the intensity values collected along the radar profiles shown in Fig. 1b. The procedure and the parameters used are reported in Lauro et al.[37].

To generate the map reported in Fig.5 we first applied a mesh refinement technique to obtain pixels containing about 100 samples; all pixels having fewer samples were discarded. For each pixel we computed a probability density function of the basal permittivity and assigned the median value of such distribution to the pixel coordinates in the scatter map (Supplementary Fig. 6). Our analysis was focused on the area having the highest pixel density (white dashed box in Supplementary Fig. 6) which was interpolated to generate a contour map (Fig. 5). Nevertheless, the areas with lower pixel densities (outside the white dashed box) are very consistent (spatial continuity) and characterized by low permittivity values (dry rocks) except for a small area in the North-East, where the permittivity values are higher. This latter area was not included in the analysis given the low MARSIS coverage and the lack of crossing orbits.

Note that neither the surface nor basal roughness were accounted for in the inversion procedure. In the area, the surface is very smooth at the MARSIS scale[63] whereas the basal roughness is not well constrained. Therefore, the basal permittivity computed with the inversion could be, in principle, underestimated[64]; to mitigate such an effect the permittivity values associated with acuity values lower than 0.5 (half of the maximum acuity[34]) were filtered off the map (Fig.5). Note that the subsurface and surface acuity values for the main body of liquid water are similar (Fig.3b) and the acuity values computed for the other patches of water are only slightly lower, suggesting that, if present, the underestimation of the permittivity should not be significant.